\documentstyle[epsf,twoside,fleqn,espcrc2]{article}
\def\gevc {\ifmmode {\rm GeV/c} \else ${\rm GeV/c}$\fi}
\def\mevc {\ifmmode {\rm MeV/c} \else ${\rm MeV/c}$\fi}
\def\gevcc {\ifmmode {\rm GeV/c^2} \else ${\rm GeV/c^2}$\fi}
\def\mevcc {\ifmmode {\rm MeV/c^2} \else ${\rm MeV/c^2}$\fi}
\def\invpb {\ifmmode {\rm pb^{-1}} \else ${\rm pb^{-1}}$\fi}
\def\invfb {\ifmmode {\rm fb^{-1}} \else ${\rm fb^{-1}}$\fi}
\def\tento {\ifmmode {\rm 10^{31}\ cm^{-2} sec^{-1}} \else ${\rm 10^{31}\
cm^{-2} sec^{-1}}$\fi}
\def\tentt {\ifmmode {\rm 10^{32}\ cm^{-2} sec^{-1}} \else ${\rm 10^{32}\
cm^{-2} sec^{-1}}$\fi}
\def\Vtdts{\ifmmode {|{V_{td}}/{V_{ts}}|} \else {$|{V_{td}}/{V_{ts}}|$}\fi}
\def\abs#1{\left| #1\right|}

\title{$B$ Physics with the CDF Run II Upgrade
\hfill FERMILAB-Conf-95/408-E
}
\author{Fritz DeJongh\address{Fermilab M.S. 318, P.O. Box 500,
Batavia IL 60510}}
\begin{document}
\begin{abstract}
We summarize Run I results relevant to an analysis of
the CP asymmetry in $B\to J/\psi K_s$, the CDF upgrade plans for Run II,
and some of the main $B$ physics goals related to the exploration of the
origin of CP violation.
\end{abstract}
\maketitle
\section{INTRODUCTION}

During the Run I data taking period, from 1992 through 1995, CDF
has acquired 110 \invpb\ of $p \bar{p}$ collisions at a center of mass
energy of 1800 GeV.  This data has provided many results on $B$
physics~\cite{Meschi}, and provides a basis for extrapolating to
Run II, which is scheduled to start in 1999 after major upgrades to both the
accelerator and detector.

We present herein a summary of Run I results relevant to an analysis of
the CP asymmetry in $B\to J/\psi K_s$, the CDF upgrade plans for Run II,
and some of the main $B$ physics goals related to the exploration of the
origin of CP violation.

\section{TAGGED $B\to J/\psi K_s$ IN RUN I}
\label{section_runi}

\begin{figure}[h]
\epsfysize=3.0in
\epsffile{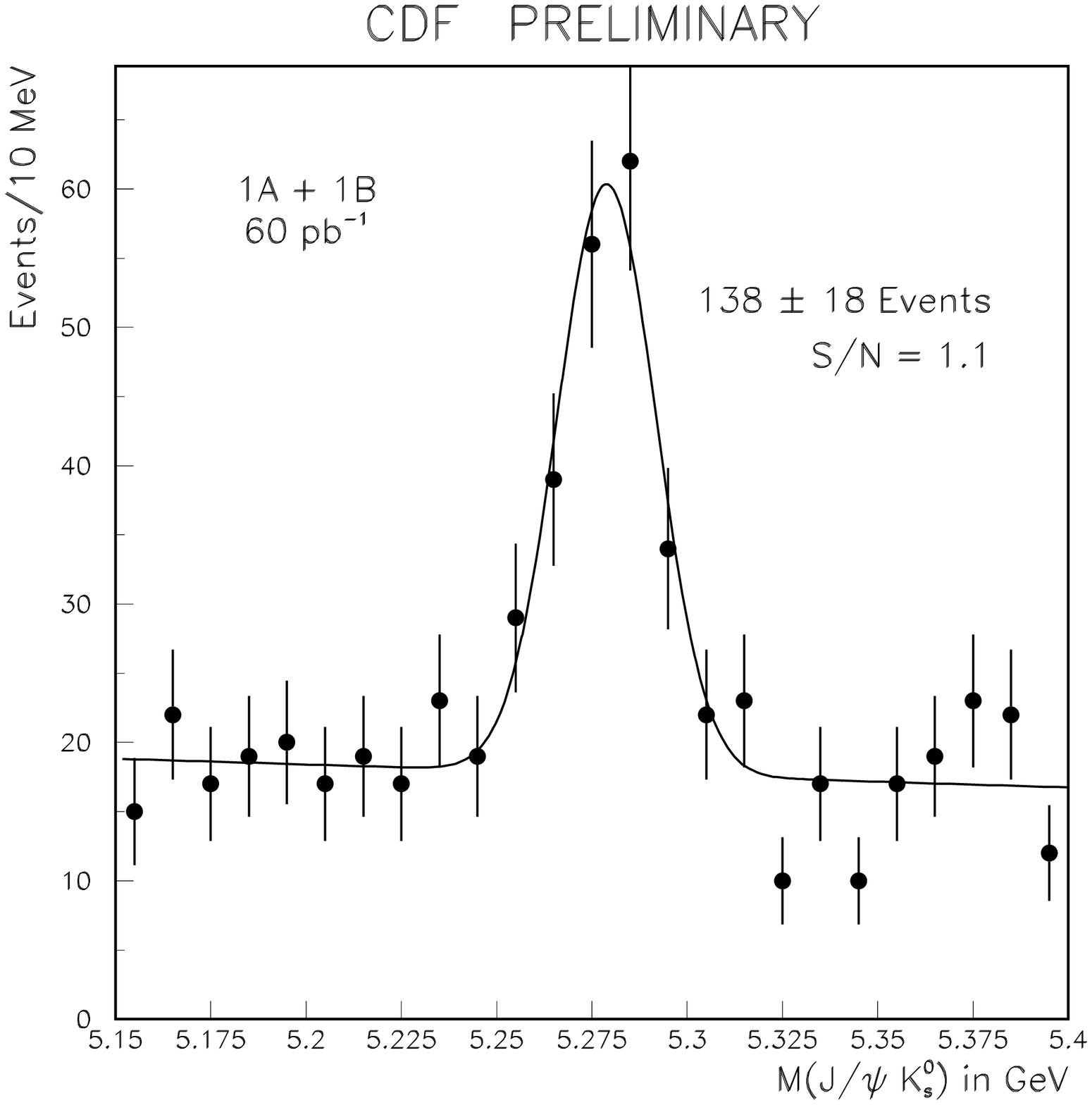}
\caption[ $B_d \to J/\psi K_s$ signal from the CDF experiment ]
{$B_d \to J/\psi K_s$ signal.}
\label{fig:b01_jpsi_kshort}
\end{figure}

In the first 60 pb$^{-1}$ of Run I, as shown in
Figure~\ref{fig:b01_jpsi_kshort},
we have observed 140 $B^0 \to J/\psi K_s$
events with signal-to-noise better than 1:1.  We obtained this sample with
a dimuon trigger that required both muons to have transverse
momentum ($P_T$) greater than 2.0 GeV.
To obtain the CP asymmetry we must tag the flavor of the $B$ meson at the time
at which it was produced.  Work is under way to use a combination
of Run I data and Monte Carlo to establish the
``effective tagging efficiency'' $\epsilon (1-2w)^2$,
where $\epsilon$ is the tagging efficiency and $w$ is the
mistag fraction, for a variety of algorithms.
We currently have results~\cite{Manfred} for two methods, Jet Charge and
Muon tagging, for a total effective tagging efficiency of $\approx 2\%$.
These results indicate that an order of magnitude improvement in the
statstical uncertainty on the CP asymmetry will lead to a competitive
measurement in Run II.

\section{ACCELERATOR IMPROVEMENTS FOR RUN II}

A project called Fermi III is underway~\cite{fermi3} to upgrade the Fermilab
accelerator complex to produce an order of magnitude higher luminosity
in the Tevatron.
The luminosity in Run I was limited by the antiproton current.
The largest component of Fermi III is to replace the Main Ring, which is
housed in the same tunnel as the Tevatron and provides the acceleration
stage just prior to the Tevatron, with the Main Injector, which will
be housed in a separate and new tunnel.  The Main Injector will provide
for higher proton intensity onto the antiproton production target,
and larger aperture for antiproton transfer into the Tevatron.
Combined with improvements to the antiproton cooling system, the
antiproton stacking rate will increase by over a factor of three
to $17 \times 10^{10}$ per hour.

\begin{figure}[h]
\epsfysize 8cm
\epsffile{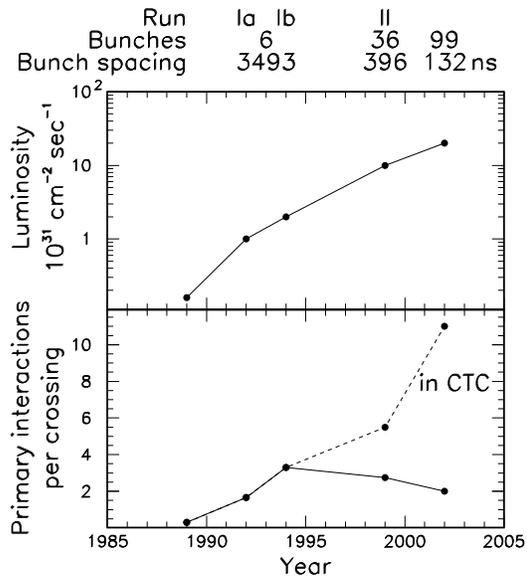}
\caption{The Tevatron schedule. As the luminosity increases, the number of
bunches increases, and the
number of primary interactions per bunch crossing remains at Run I levels.
The current Central Tracking Chamber (CTC) has a drift time
of 750 ns, and thus at any given time is occupied by events from more than
one bunch crossing.  }
\label{fig:schedule}
\end{figure}

The Tevatron schedule and some basic parameters are shown in
Figure~\ref{fig:schedule}.
Our physics projections for Run II assume 2 \invfb\ of integrated luminosity.

\section{DETECTOR IMPROVEMENTS FOR RUN II}

The CDF detector is being upgraded to handle an order of magnitude higher
luminosity, and 132 ns bunch spacing~\cite{PAC}.
The main goal is to maintain detector
occupancies at Run I levels,
although many of these upgrades also provide for qualitatively
improved detector capabilities.

\subsection{Tracking Upgrades}

The efficiency of the current tracking system would be significantly
degraded at luminosities of $\tentt$:  Primary track efficiency would
drop by 10\%, and $K_s$ efficiency would drop by 60\%.  A three part
tracking upgrade is being planned to recover this efficiency:
\begin{itemize}
\item A Central Straw Tracker (CST) will consist of 4 axial and 4 stereo
superlayers of 8 to 12 straws each at radii of 50 to 140 cm.
\item An Intermediate Scintillating Fiber Tracker (IFT) will consist
of six stereo and six axial layers of 500 micron diameter scintillating
fibers read out by VLPCs.
\item A new Silicon Vertex Detector (SVX II).  The SVX II will consist
of 5 layers of double sided silicon from radii of 2.9 to 10 cm, arranged
in 5 axial layers, 2 small angle ($1.2^\circ$) stereo layers, and 3
$90^\circ$ stereo layers.
\end{itemize}

Since each of these detectors can be read out in less than 132 ns,
occupancies will be held to Run I levels up to luminosities of
$3 \times \tentt$ at 132 ns bunch spacing.  Furthermore, each
detector is potentially a stand-alone 3D tracker, providing for greater
redundancy in pattern recognition.

In addition to maintaining efficiency, these upgrades provide for new
tracking capabilities:  Precision vertexing in 3 dimensions, tracking
to $\abs{\eta} <2$ and tracking down to $p_T > 100 \mevcc$.

\subsection{Time of Flight}

We are planning for a Time of Flight system consisting of 3 m long
$4 \times 4$ cm scintillator blocks placed at a radius of 1.4 m
(inside the solenoid) and
read out with mesh dynode photomultiplier tubes.
The 4 cm width results in less than 20\% confusion from
multiple hits and other noise sources.
We expect 100 to 125 ps time resolution, for $>2\sigma$ $K/\pi$
separation in the momentum range from 0.3 to 1.6 \gevc.
This momentum range includes 55\% of kaons potentially useful for
flavor tagging.

\subsection{Trigger and DAQ system}

\begin{figure}[h]
\epsfysize 9cm
\epsffile{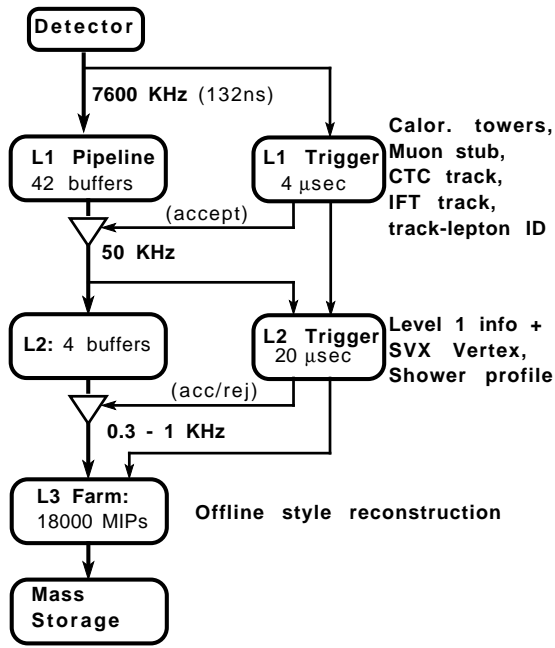}
\caption{Trigger and data acquisition flow. }
\label{fig:trigger}
\end{figure}

CDF is planning for a trigger and DAQ upgrade to allow for higher
data rates while increasing the sophistication of the trigger decision, as
summarized in Figure~\ref{fig:trigger}.  Data is stored in a 42 cell
pipeline while awaiting the Level 1 trigger decision, and can be transferred
to Level 2 without halting Level 1.  Information available for the Level 1
decision will include calorimetry clusters, CST tracks
with $p_T > 1.5\ \gevcc$,
and electron and muon identification.  At Level 2, SVX II information will
also be available, and DEC Alpha based processors allow for programmable
algorithms.  A commercial switch will be used to assemble events into
the Level 3 processors where a full event reconstruction will be performed
for the final trigger decision.

\section{$B$ PHYSICS EXPECTATIONS FOR RUN II}

The challenge for $B$ physics in Run II is to develop efficient
trigger algorithms for key final states, and efficient flavor
tagging algorithms.  In this section, we discuss possibilities for
flavor tagging, measurements of $\sin(2\beta)$, $\sin(2\alpha)$,
$B_s$ mixing, and the observation of rare decay modes.
Topics not discussed here include~\cite{pacp}:  The CP angle
$\gamma$~\cite{Snowmass_gamma}, study of exclusive $b\to u$ semileptonic
decays, measurement of $V_{cb}$ in semileptonic $\Lambda_b$ decays,
and $B_c$ spectroscopy.

\subsection{Flavor tagging}

As discussed in Section~\ref{section_runi} we
currently have results for two flavor tagging methods, jet charge and
muon tagging,
for a total effective tagging efficiency of almost 2\%.
Our goal for Run II is to attain the following
effective tagging efficiencies from various algorithms:
\begin{itemize}
\item 2\% from lepton tagging, using electrons as well as muons,
      and additional coverage planned for lepton identification in Run II.
\item 2\% from same-side tagging, which exploits the charge
      correlation of the pions produced in the fragmentation process along
      with the $B$ meson~\cite{Rosner}.  For the $B_s$ case, the charge
      correlation of kaons identified in the time-of-flight system
      may result in a tagging efficiency of 5\%.
\item 3\% from kaon tagging, using tracks with high impact parameter
      identified as kaons in the time of flight system.
\item 4\% from a jet charge algorithm expoiting
      3D vertexing information from the SVX II, and stand-alone tracking
      information from
      the SVX II and Intermediate Fiber Tracker.
\end{itemize}
While work is in progress to evaluate all these algorithms using current
data, for now we assume an 8\% flavor tagging efficiency for $B_d$ mesons,
and 11\% for $B_s$ mesons.

\subsection{CP Asymmetry in $B_d \to J/\psi K_s$: $\sin(2 \beta)$}
\label{bsection_sin2b}

With a large branching ratio and distinctive trigger signature,
the decay mode $B_d \to J/\psi K_s$ is the leading candidate for the
initial observation of CP violation in the $B$ system.  Furthermore, the
extraction of $\sin(2 \beta)$ from this asymmetry is essentially free
of hadronic uncertainties.
The current Standard Model predictions for $\sin(2\beta)$ are
$\sin(2\beta) > 0.17$~\cite{Ali-London} and
$\sin(2\beta) = 0.65 \pm 0.12$~\cite{ciuchini}.

As discussed in Section~\ref{section_runi}, CDF in Run I has
reconstructed $\approx2$ $J/\psi K_s$ events per \invpb, using a
dimuon trigger with a $p_T$ threshold of 2.0~\gevc~on each muon.
Improvements for Run II include lowering the
trigger $p_T$ threshold to 1.5~\gevc~per muon, improving the muon
coverage, and using the channel $J/\psi \to e^+ e^-$.
Our goal for Run II is to reconstruct 10 $B^0 \to J/\psi K_s$
events per \invpb, for a yield of 20,000 events assuming 2 \invfb.
We also expect to achieve much improved signal-to-noise
by using the improved capability and coverage of the SVX II, but have
conservatively assumed $S/N=$2:1.  Assuming an effective tagging
efficiency of 8\%, we find an uncertainty on $\sin(2\beta)$
of $0.07$.

In addition to the above expectation of 20,000 $B^0 \to J/\psi K_s$
events in
$2\  {\rm fb}^{-1}$, the $B^0 \to J/\psi K_s$ yield can increase by employing
(a) the increased tracking coverage and (b) new ways of triggering,
such as requiring
one lepton and one additional track with large impact parameter.
While speculative, an additional factor of four
or more in the number of reconstructed $B^0 \to J/\psi K_s$ events may
be possible.

\subsection{CP Asymmetry in $B_s \to J/\psi\phi$}
\label{section_psiphi}

Within the Standard Model, the
the CP asymmetry in $B_d \to J/\psi K_s$ measures the weak phase
of the CKM matrix element $V_{td}$, while the
CP asymmetry in $B_s \to J/\psi\phi$ measures the weak phase of the
CKM matrix element $V_{ts}$, which is expected to be very small.
As emphasized by Y. Nir~\cite{YNir}, and Helen Quinn at this workshop, the
channel $B_s \to J/\psi\phi$ may therefore provide a signature for a source of
CP violation beyond the Standard Model.
With the same trigger improvements as for $B_d \to J/\psi K_s$,
we expect 12000 $B_s \to J/\psi\phi$ events in Run II.

\begin{figure}[h]
\epsfysize=3.0in
\epsffile{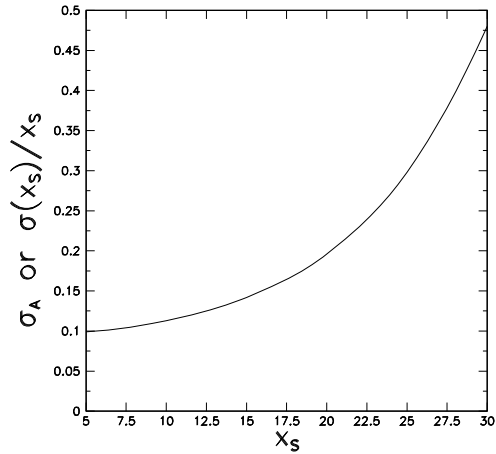}
\caption{The uncertainty on the CP asymmetry for $B_s \to J/\psi \phi$
(or the relative uncertainty on $x_s$) as
a function of the $B_s$ mixing parameter $x_s$}
\label{acp_psiphi}
\end{figure}

The magnitude of a CP asymmetry in $B_s \to J/\psi\phi$ decays would be
modulated by the frequency of $B_s$ oscillations.  Thus, for a meaningful
limit, we must be able to resolve $B_s$ oscillations.  If we neglect
resolution effects, we can expect a precision on the asymmetry of $\pm 0.09$.
However, resolution effects smear the oscillations and
produce an additional dilution.  Our experience in Run I shows that
if we determine the primary vertex event-by-event, the proper lifetime
resolution for fully reconstructed $B$ decays is
$\approx 30 \mu$m.
Figure~\ref{acp_psiphi} shows our expected
precision on the asymmetry as a function of $x_s$.
There will be an additional dilution if the $J/\psi \phi$ final
state is not a pure CP eigenstate.

\subsection{CP Asymmetry in $B^0 \to \pi^+\pi^-$: $\sin(2 \alpha)$}

A measurement of $\sin(2 \alpha)$ in conjunction with
$\sin(2 \beta)$ provides powerful constraints on the
unitarity triangle~\cite{Lauten}.
The greatest challenge in this measurement is the trigger requirement
at a luminosity of $1 \times 10^{32} {\rm cm}^{-2}{\rm sec}^{-1}$.
Our plan (described in detail in~\cite{Snowmass_pipi}) consists of
\begin{enumerate}
\item At Level-1: Require two tracks with $P_T>2$ \gevc,
imposing $\Delta\phi$ cuts on opposite sign track pairs.
\item At Level-2: Require an impact parameter
 $> 100 ~\mu$m for each track.
\item At Level-3: Use the full event information for the final decision.
\end{enumerate}
After additional analysis
requirements we expect $\approx 5$ $B^0 \to \pi^+\pi^-$ events
per pb$^{-1}$.  Due to the impact parameter cuts, the
proper lifetime distribution starts at $\approx 1.5$ lifetimes,
and  therefore
the dilution of the CP asymmetry due to mixing of the signal $B$ before
it decays will be 0.82, rather than 0.47 as we assumed for $\sin(2\beta)$.

\begin{figure}[h]
\epsfysize=3.0in
\epsffile{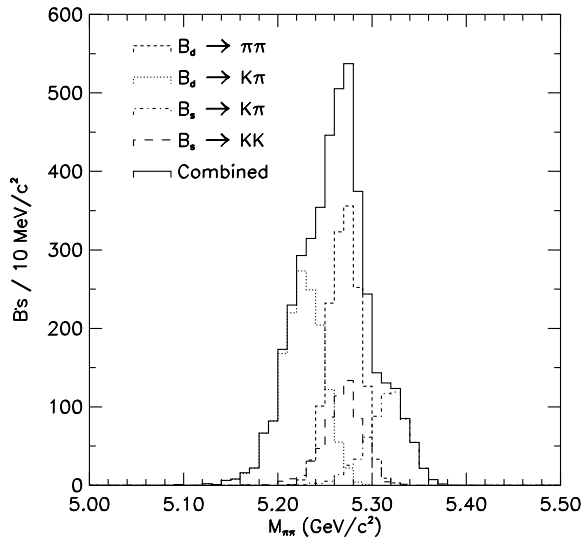}
\caption{Mass distribution for the combination
$B^0 \to \pi^+\pi^-$, $B_d \to K^+ \pi^-$,
$B_s \to K^-\pi^+$ and $B_s \to K^+ K^-$
assuming a pion mass assignment for all tracks.}
\label{fig:b04_pipimass}
\end{figure}

To measure the CP asymmetry in $B^0 \to \pi^+\pi^-$ events one needs to
determine the fraction of the signal from $B_d \to K^+ \pi^-$,
$B_s \to K^- \pi^+$ and
$B_s \to K^-  K^+$ decays.  This can be done using invariant mass and
$dE/dx$ distributions in the high statistics untagged sample.
Figure~\ref{fig:b04_pipimass} displays the
expected mass distribution for the combination of the above four signals,
assuming a pion mass assignment for all tracks~\cite{Snowmass_pipi}.
The $B_d \to \pi^+ \pi^-$
and $B_d \to K^+ \pi^-$ peaks are separated by 40 \mevcc, while we
expect a mass resolution of $\approx 28 \mevcc$.  As in Run I, we
also expect $K/\pi$ separation from $dE/dx$ in the CST of
better than $1\sigma$.  Any CP asymmetry in the $K\pi$ background
component can be determined from the ratio of numbers of $K^-\pi^+$
and $K^+\pi^-$ events in the untagged sample.  Any CP asymmetry in the
$K^+ K^-$ background component would be modulated by the $B_s$ mixing
frequency rather than the $B_d$ mixing frequency.  Therefore, the CP
asymmetry in the tagged sample in conjunction with a fit to the untagged
sample can yield the CP asymmetry for $B^0 \to \pi^+\pi^-$.

Another issue for this analysis is the combinatorial background under
the $B$ peak.  We can estimate this background level
detector using a sample of high $E_T$ electron triggers from Run I.
In the case that the electron results from the semileptonic decay
of a $B$ hadron, we can search for the other $B$ in the event to
decay to $\pi^+ \pi^-$.
Using standard cuts on the decay vertex and the isolation of the two-track
combination, we obtain an observed background, $N$, comparable to the
expected signal, $S$ (less than one event),
for $P_T > 4$ \gevc~on each track: $S/N \approx 1:1$.
Lowering the $P_T$ threshold to 2 \gevc~will allow us to double our efficiency.
We expect to do this with the Run II detector while maintaining
$S/N$ better than 1:1
by exploiting the 3D information from the SVX II and optimizing cuts.

The final issue related to the extraction of the angle $\alpha$ from
the measured CP asymmetry in $B_d \to \pi^+\pi^-$ is the extraction of
possible penguin contributions in addition to the tree diagram which is
expected to dominate this decay mode.  We can estimate this
penguin contamination, and thus extract $\alpha$, from a combination
of experimental measurements and theoretical inputs.  In particular,
a time-dependent analysis yields a
measurement of the amplitude as well as the phase of the CP asymmetry,
which oscillates with the mixing frequency.
This latter phase would be zero in the absence of a penguin contribution.
In addition, we use the average branching ratio
$\left(Br(B^0 \to \pi^+\pi^-)+Br(\bar B^0 \to \pi^+\pi^-)\right)/2$.
This quantity can be extracted from
untagged $B_d \to \pi^+\pi^-$ decays and will therefore have a very small
error.   Other ingredients are the value of $Br(B \to \pi \ell \nu)$
and some
theoretical input such as the magnitude of the tree diagram given
$Br(B_d \to \pi^- \ell \nu)$.
As an example, if the penguin amplitude, $A_p$,
is small compared to the tree amplitude, $A_t$,
(say, $A_p/A_t = 0.07$ as predicted by Deshpande {\it et. al.}~\cite{DH1})
the extraction of $\alpha$ is relatively easy, and
the theoretical constraints can be relatively crude.
If $A_p/A_t \approx 0.2$, this becomes more challenging, but feasible.
A detailed analysis can be found in reference~\cite{Fritz_Paris}.

In conclusion, assuming a flavor-tagging efficiency of $8\%$ as for the
$J/\psi K_s$ case, and a conservative $S/N = 1/4$, we expect an
overall uncertainty on $\sin(2 \alpha)$ of $\pm 0.10$.

\subsection{Measurement of \Vtdts}

	The CDF $B$ physics goals for Run II include observation of the
time dependence of $B_s$ and $B_d$ mixing, obervation of exclusive
radiative penguin decays, and the observation of a lifetime difference
$\delta \Gamma$
between the CP eigenstates of the $B_s$ meson.
	The ratio of $B_d$ to $B_s$ mixing parameters, $x_d/x_s$,
is proportional to $\Vtdts^2$.  The matrix elements for $B_d$ and $B_s$
mixing are related by SU(3), allowing the cancellation of many
theoretical uncertainties in the ratio.
	Similarly, in the absence of long distance effects, the
ratio of decay rates
$B(B\to\rho\gamma)/B(B\to K^*\gamma)$ is proportional to $\Vtdts^2$.
Again, since these final states are related by SU(3), many
theoretical uncertainties cancel in the ratio.

	A smaller value of \Vtdts\ implies a larger value of $x_s$,
and a smaller rate for $B\to\rho\gamma$, and therefore both of these
measurements become more difficult.  However, the lifetime difference
$\Delta \Gamma_s$ is proportional to $x_s$, and therefore this
measurement becomes easier.  Although the theoretical uncertainties
on $\Delta \Gamma_s$ are larger, the combination of the three types of
measurements
discussed in this section should allow CDF to measure \Vtdts\
over the full range permitted by the standard model.

\subsubsection{$B_s$ Mixing}

ALEPH has shown that $x_s > 8.8$~\cite{aleph}, implying that
$x_s$ must be measured by fitting the time-dependent
oscillation of the $B_s$.
For the Run Ia $B_s$ lifetime measurement~\cite{Bs-ctau}, CDF reconstructed
70 $B_s\rightarrow D_s \ell \nu$ events.  For Run II, triggering and
reconstruction of this channel with very high statistics is straightforward,
and we expect $\approx 10^5$ events.  Due to the unreconstructed neutrino,
knowledge of the $B$ momentum limits
the measurement to values of $x_s < \sim 11$.
However, improvements in the analysis technique may result in an
improved momentum resolution.  For example, 3D vertexing
allows a determination of the four-momentum of the missing neutrino,
although with a quadratic ambiguity.

For fully reconstructed decays, the $x_s$ reach is limited by
vertexing resolution, as discussed in section~\ref{section_psiphi}.
In order to determine the flavor of the $B_s$ at the time of the decay
this measurement requires events of the type $B_s \rightarrow D_s n\pi$.
The challenge for CDF is to trigger on, and isolate from background,
signals of this type.  We note that the presence of
a time-of-flight system in CDF should significantly improve the
reconstruction efficiency by allowing efficient selection of kaons
and rejection of pions  at low $P_T$, where the backgrounds are largest.

One strategy is to
trigger on a single lepton ($e$ or $\mu$), which will serve as the
flavor tag, and then reconstruct $B_s$ decays in this sample~\cite{Snowmass}.
For a 6 GeV lepton threshold in Run II, there will be
$\sim 2 \times 10^3$ $B_s$ mesons that have decayed
within the CDF fiducial volume to the modes
$B_s \rightarrow D_s + \pi$ and $B_s \rightarrow D_s + 3\pi$
with $D_s\rightarrow \phi \pi$
or $D_s \rightarrow K^{*\pm} K^\mp$.
It is not yet known how many of these may be reconstructed with good
signal-to-noise.
It is likely that the lepton trigger threshold will be lower, and also that
some of the decay products of the $B_s$ will be included in the trigger
requirement as well.

Another strategy is to use a fully hadronic trigger,
as for $B \to \pi^+\pi^-$, in which
case all tagging techniques may be applied.
The final states
of the $B_s$ we are trying to reconstruct are produced an order of
magnitude more frequently than $B \to \pi^+\pi^-$.
More work is needed to design such a trigger;  one possibility
is a two-track trigger optimized for $\phi \to K^+ K^-$.

Although we do not have a solid projection of how many events we will
reconstruct, we show in Figure~\ref{acp_psiphi} our precision on $x_s$ if
we succeed in reconstructing 2000 events from fully hadronic triggers,
with an effective tagging efficiency of 11\%, or equivalently,
800 events recontructed in events with a lepton trigger.

%\begin{figure}[h]
%\epsfysize=3.0in
%\epsffile{sig_xs.eps}
%\caption{Relative uncertainty on $x_s$ for fully reconstructed
%$B_s$ decays.}
%\label{sig_xs}
%\end{figure}

\subsubsection{Radiative $B$ Decays}

Measurements of radiative $B$ decays at CLEO sets an upper bound
on \Vtdts of 0.76~\cite{CLEO-rare}.
CDF has already installed a trigger to collect radiative penguin decays.
The limited
bandwidth available in the current trigger and data acquisition system
require the trigger to have quite high thresholds (10 GeV
photon plus two 2 GeV tracks).  The expected yield with this
trigger is $\approx 20$ $\gamma K^*$ events per 100 pb$^{-1}$.
In Run II,
we expect to lower the photon $E_t$ threshold to 5 GeV and the track
$P_t$ threshold to 1.5 GeV, with a resulting yield of $\sim 135$ events
per 100 pb$^{-1}$ or $\sim 2700$ for 2 fb$^{-1}$.

\begin{figure}[ht]
\epsfysize=3.0in
\epsffile{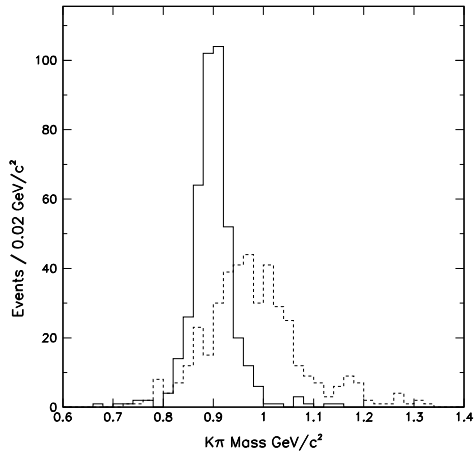}
\caption{A Monte Carlo simulation for $B\rightarrow K^*\gamma$ (solid) and
$B\rightarrow \rho \gamma$ (dashed) reconstructed
as $B\rightarrow K^* \gamma$.}
\label{fig-peng}
\end{figure}

The mass resolution of the reconstructed $B$
is dominated by the resolution on the photon energy and is $\sim 140$
MeV.  We have studied our ability to reject combinatorial background
using Run 1A photon data and have studied with Monte Carlo the
discrimination against $B\rightarrow K^* \pi^0$ and
$\rho \pi^0$ and from higher multiplicity penguin decays.
These backgrounds are manageable.  The mass resolution is not adequate
to separate $\gamma \rho$ from $\gamma K^*$ on an event-by-event
basis (see Figure~\ref{fig-peng}); however, a statistical separation is
possible.  In addition, the CTC dE/dx system should
provide 1$\sigma$ $K$-$\pi$ separation in the momentum range of
interest.

These radiative $B$ decays can also be observed using converted photons.
The probability for a photon to convert ($\sim 5\%$)
will be offset by a lower photon $E_t$ threshold.  Also, the mass
resolution is $\sim 5$ times better than for the signals with
unconverted photons, allowing a cleaner separation between
$B\to\gamma K^*$ and $B\to\gamma\rho$.

At the Tevatron it is possible to study $B_s$ penguin
decays as well.  Information on \Vtdts\ can be obtained in the same
manner as above from studying the ratio of
$B(B_s\rightarrow \gamma K^*)/B(B_s \rightarrow \gamma \phi)$.
The size of the $B_s$ penguin sample is expected to be 1/2 to 1/3
the size of the $B_d$ sample.
Comparison of the two results
would help constrain the size of the long distance contributions to
the decays.

\subsubsection{$\Delta \Gamma_s/\Gamma_s$}

Browder {\it et al.}~\cite{Browder}
show that if $x_s=15$, a 7\%\ difference in lifetime is
expected.  Several techniques can be used to
determine $\Delta \Gamma_{B_s}$~\cite{Dunietz}.
The statistical uncertainty on the $B_s$ lifetime from semileptonic
$B$ decays in Run II will be well below $1\%$.
With this constraint, the decay mode
$B_s \rightarrow J/\psi \phi$
can be decomposed into its two
CP components (via a helicity analysis) fitting a separate
lifetime for each component (if this final state is a pure CP eigenstate,
its lifetime can simply be compared to the average $B_s$ lifetime).
Using Run Ia data, CDF has measured the helicity structure of the decays
$B\rightarrow J/\psi K^*$ and $B\rightarrow J/\psi \phi$~\cite{randy}.
In Run II, the $B_s \rightarrow J/\psi \phi$
helicity structure should be known to about 1\%,
and the lifetime difference should be determined
to 2-3\%.

\subsection{Rare $B$ decays}

In Run I, CDF has performed a search
for the decay modes $B^\pm \to \mu^+\mu^- K^\pm,
B^0 \to \mu^+\mu^- K^{*o}$ and $B_{d,s} \to \mu^+\mu^-$~\cite{Meschi}.
Assuming the Standard Model Branching ratios~\cite{Ali-ECFA}
for $B^+ \to \mu^+\mu^- K^+$ and $B^0 \to \mu^+\mu^- K^{*o}$,
we expect in Run II
$\approx 400$ $B^+ \to \mu^+\mu^- K^+$ and
$\approx 650$ $B^0 \to \mu^+\mu^- K^{*o}$ events.  This will enable us to
study both (a) the invariant mass distribution of the dimuon pair
and (b) the forward-backward charge asymmetry in the decay.  Both
of these distributions are sensitive to physics beyond the Standard
Model, e.g. the presence of a charged Higgs or charginos~\cite{wyler},
\cite{ali-model-indep}.

We also expect to oberve the decays $B^\pm\to e^+ e^- K^\pm$ and
$B^0 \to e^+ e^- K^{*0}$.
The decays $B^\pm \to\ell^+\ell^+\pi^\pm$ and
$B^0\to\ell^+\ell^-\rho^0$ are suppressed by
an order of magnitude, but will be observable if we can achieve a high
enough level of signal-to-noise.  An observation of these decay modes
would provide another determination of $|V_{td}/V_{ts}|$.

\section{CONCLUSIONS}

The CDF Run I has provided much experience in doing $B$ physics at a
hadron collider, including the reconstruction of $B \to J/\psi K_s$, and
flavor tagging.  This experience indicates that with an order of
magnitude improvement in statistical sensitivity, we can obtain
a competitive measurement of $\sin(2\beta)$ in Run II.  We expect to obtain
this factor from
accelerator upgrades, which
will provide an order of magnitude more integrated
luminosity, and detector upgrades, which will improve trigger and flavor
tagging efficiency.
Other $B$ physics goals include the measurement of the CP asymmetry in
$B\to \pi^+\pi^-$, the observation of $B_s$ mixing, and high statistics
observations of certain rare decay modes.  Many fundamental measurements
involve the $B_s$, $B_c$, or $\Lambda_b$, and are thus complementary
to $B$ physics programs at the $\Upsilon(4S)$.

\end{document}